\documentclass[english,aps,prx,twocolumn,amsmath,amssymb,showpacs,superscriptaddress,notitlepage,longbibliography]{revtex4-1}
\usepackage{bm}
\usepackage{amsmath}
\usepackage{amssymb}
\usepackage{mathdots}
\usepackage{graphicx}
\usepackage{hyperref}
\usepackage{xcolor}

\usepackage{braket}
\usepackage{bm}

\begin{document}
	\title{Geometric and conventional contributions of superconducting diode effect: Application to flat-band systems}
            
	\author{Jin-Xin Hu}\thanks{jhuphy@ust.hk}
	\affiliation{Division of Physics and Applied Physics, School of Physical and Mathematical Sciences, Nanyang Technological University, Singapore 637371} 
	\affiliation{Department of Physics, Hong Kong University of Science and Technology, Clear Water Bay, Hong Kong, China}
        \author{Shuai A. Chen}
\affiliation{Max Planck Institute for the Physics of Complex Systems, N\"{o}thnitzer Stra{\ss}e 38, Dresden 01187, Germany}
        \author{K. T. Law}\thanks{phlaw@ust.hk}
        \affiliation{Department of Physics, Hong Kong University of Science and Technology, Clear Water Bay, Hong Kong, China}
        
\begin{abstract}	
Nonreciprocal critical supercurrents give rise to the superconducting diode effect (SDE) in noncentrosymmetric superconductors when time-reversal symmetry is broken. In this paper, we investigate the SDE in
superconductors with vanishing spin-orbit coupling but featuring narrow bands near the Fermi energy—a characteristic particularly relevant to moir\'{e} heterostructures, such as twisted bilayer graphene. Using phenomenological Ginzburg-Landau theory and self-consistent mean-field approaches, we analyze the contributions to the SDE from both conventional band dispersion and quantum geometry. While the conventional SDE arises from the asymmetric Fermi surface, we demonstrate that the quantum metric dipole generates a band quantum-geometric contribution to the SDE, even in systems with symmetric single-particle dispersion. Notably, in the flat-band limit, where the attractive interaction strength significantly exceeds the bandwidth, the contributions from quantum geometry to the supercurrent and diode effect become dominant. Our paper elucidates the conventional and quantum-geometric origins of superconducting nonreciprocity and explores their implications for flat-band superconductors.
\end{abstract}
\maketitle

\section{Introduction}
The nonreciprocal responses have garnered significant attention due to its fundamental importance in quantum materials and devices~\cite{tokura2018nonreciprocal,wakatsuki2017nonreciprocal,nagaosa2023nonreciprocal}. In superconductors, this phenomenon is exemplified by the superconducting diode effect (SDE)~\cite{wakatsuki2018nonreciprocal,jiang2022superconducting,nadeem2023superconducting}, which has been observed in bulk superconductors~\cite{ando2020observation,lyu2021superconducting,lin2022zero,narita2022field,hou2023ubiquitous} and Josephson junctions~\cite{wu2022field,pal2022josephson,diez2023symmetry,trahms2023diode}. The SDE is characterized by the asymmetry in critical or depairing supercurrents between the right and left directions. It is widely accepted that the breaking of both inversion ($\mathcal{P}$) and time-reversal ($\mathcal{T}$) symmetries is essential for the emergence of the SDE. To understand the SDE, numerous theoretical mechanisms have been proposed~\cite{zinkl2022symmetry,yuan2022supercurrent,he2022phenomenological,daido2022intrinsic,ilic2022theory,daido2022superconducting,zhai2022prediction,scammell2022theory,legg2022superconducting,davydova2022universal,zhang2022general,xie2023orbital,hu2023josephson,he2023supercurrent,banerjee2024enhanced}. These studies primarily focus on the deformation of the Fermi surface, particularly changes in band dispersion caused by effects such as the Zeeman effect~\cite{yuan2022supercurrent,he2022phenomenological,daido2022intrinsic} or trigonal warping~\cite{scammell2022theory,hu2023josephson}.

\begin{figure}
		\centering
		\includegraphics[width=1.0\linewidth]{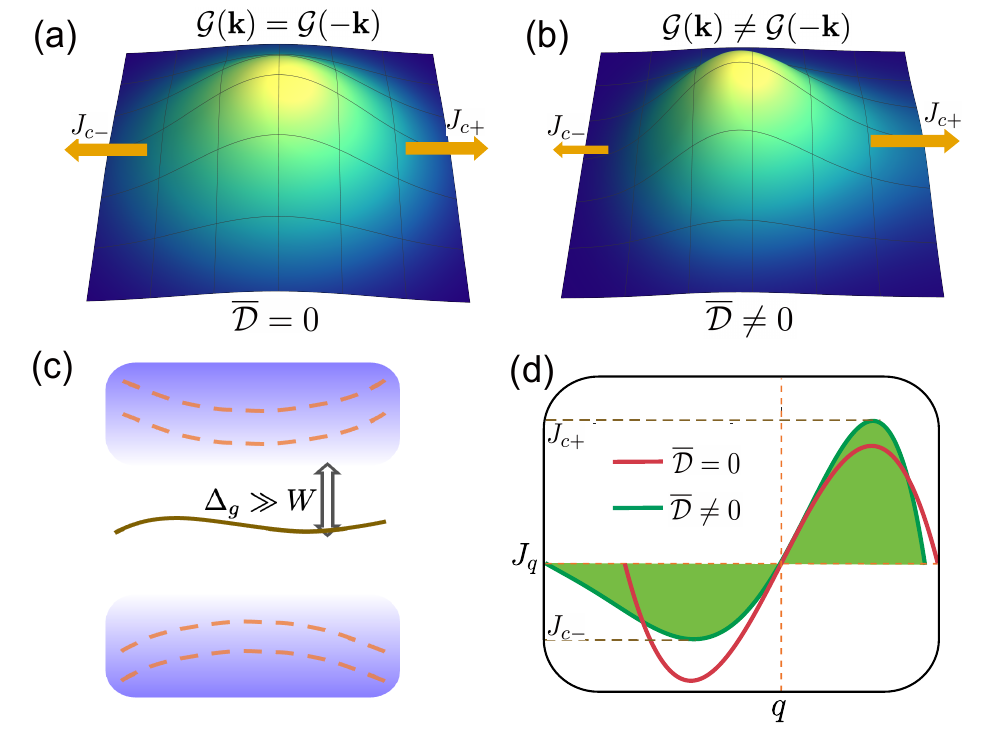}
		\caption{(a) The profile of quantum metric $\mathcal{G}$ in the momentum space. The depairing current is reciprocal with $|J_{c+}|=|J_{c-}|$. In this case, the preservation of $\mathcal{P}$ enforces that the $\mathcal{G}(\bm{k})$ satisfies $\mathcal{G}(\bm{k})=\mathcal{G}(-\bm{k})$, and the average of the quantum metric dipole $\overline{\mathcal{D}}$ is 0. (b) In the case of $\mathcal{G}(\bm{k})\neq \mathcal{G}(-\bm{k})$, the presence of a non-zero $\overline{\mathcal{D}}$ results in $|J_{c+}|\neq|J_{c-}|$. The $\mathcal{T}$ symmetry is already broken by valley polarization. (c) A multi-band system with an isolated flat band at the Fermi energy, and the band gap $\Delta_g$ separating it from other bands (depicted by the orange dashed lines) is significantly larger than the bandwidth $W$. (d) A schematic illustration of the current-$q$ relation for (a) and (b), where $q$ is the Cooper pair momentum. }
		\label{fig:fig1}
\end{figure}

Nevertheless, the role of the quantum geometric effect of wave functions, as an exclusively multi-band effect, remains relatively unexplored in shaping the SDE, particularly in the context of flat-band superconductors. In a flat-band superconductor for which the conventional BCS relation fails since the Fermi velocity vanishes, the wave-function quantum geometry can define the superfluidity and the coherence length~\cite{peotta2015superfluidity,julku2016geometric,liang2017band,verma2021optical,torma2022superconductivity,
herzog2022superfluid,torma2023essay,chen2024ginzburg,hu2025anomalous}. Thus the quantum geometric effect can be considered as an additional degree of freedom to manipulate the physical properties apart from the electronic band structures. Therefore, investigating the interplay between the quantum geometric effect and the SDE is of fundamental importance which also holds the potential for driving further experimental developments.

 The discovery of moir\'{e} superconductors such as magic angle twisted bilayer graphene provides a platform for exploring the flat-band superconductivity, and sizable supercurrents were observed in these experiments~\cite{cao2018unconventional,yankowitz2019tuning,arora2020superconductivity,oh2021evidence,
tian2023evidence}. The non-dissipative transport in the moir\'{e} superconductors is associated with the finite superfluid weight, primarily governed by quantum geometry~\cite{hu2019geometric,julku2020superfluid,xie2020topology,torma2022superconductivity}. On the other hand, the observation of the SDE in twisted trilayer graphene with vanishing spin-orbital coupling is particularly interesting \cite{lin2022zero}. This phenomenon is notable for its manifestation in the absence of a magnetic field and the presence of narrow bands near the Fermi energy.

Motivated by the recent experimental progresses, one fundamental question arises: Can quantum geometry play any role in the nonreciprocal SDE? In this work, we unveil the SDE originating from quantum geometry of the Bloch wavefunction. Explicitly, we consider an isolated narrow band with the energy $\epsilon_{\bm{k}}$ while the wavefunction $|u_{\bm{k}}\rangle$  possesses finite quantum metric $\mathcal{G}$ defined as 
\begin{equation}
\label{eq:q_matric}
\mathcal{G}_{ab}(\bm{k})=\mathrm{Re}[\langle \partial_a u_{\bm{k}}|\partial_b u_{\bm{k}}\rangle-\langle \partial_a u_{\bm{k}}|u_{\bm{k}}\rangle\langle u_{\bm{k}}|\partial_b u_{\bm{k}}\rangle],
\end{equation}
where $\partial_a=\partial/\partial_{k_a}$ is the momentum derivative.
By establishing the Ginzburg-Landau theory around the transition region, we study the dissipationless supercurrent as well as the feasibility of SDE (see Fig.~\ref{fig:fig1}). Importantly, the geometric origin of SDE is the distribution of quantum metric dipole with the expression as 
\begin{equation}
\label{eq:q_dipole}
\mathcal{D}_a^{bc}(\bm{k})=\partial_a \mathcal{G}_{bc}(\bm{k}).
\end{equation} 
The average of quantum metric dipole $\overline{\mathcal D}_a^{bc}$ [see Eq.~~\eqref{eq:dipole_ave}] over the first Brillouin zone can be finite when the $\mathcal{P}$ symmetry is broken [see Fig.\ref{fig:fig1}~(a), (b)]. The SDE can show up when the $\mathcal T$ symmetry is also violated. By utilizing the phenomenological analysis, we calculate the Ginzburg-Landau coefficients for a Haldane-like lattice model with flattened bands which consists of finite quantum metric and a non-zero quantum metric dipole. In conjunction with the phenomenological analysis utilizing the Ginzburg-Landau theory, we calculate the nonreciprocal critical currents by a mean-field study of the lattice model and analyse the SDE from conventional band dispersion and quantum geometry. Furthermore, we discuss the potential application of this theoretical framework to moir\'e superconductors, which exhibits isolated narrow bands near the Fermi energy. The implications of our findings can provide valuable insights for both theoretical and experimental investigations.

\section{Ginzburg-Landau theory for SDE}
To elucidate the fundamental physical properties, we consider a multi-band system which possesses an isolated narrow band near Fermi energy as shown in Fig.~\ref{fig:fig1}(c). There is a large band gap $\Delta_g$ significantly exceeding the bandwidth of the isolated band $W$. Initially, we have $H=H_0+H_{\mathrm{int}}$, where the non-interacting $H_0$ describes the band structure of bare electron $a_{\sigma}(\bm{r})$. For simplicity, we focus on $s$-wave pairing and $H_{\mathrm{int}}$ involves the effective attractive interaction
\begin{equation}\label{eq:hint}
H_\mathrm{int}  =-U\int d\bm{r}a_{+}^{\dagger}(\bm{r})a_{-}^{\dagger}(\bm{r})a_{-}(\bm{r})a_{+}(\bm{r})~,
\end{equation}
where $U$ denotes attractive interaction strength, and $\sigma=\pm$ denotes the flavor (spin, valley) index. Owing to the large band gap, we make the projection onto the isolated band, which yields
\begin{equation}
a_{\sigma}(\bm{r})= \frac{1}{\sqrt{N}}\sum_{\alpha\bm{k}}e^{i\bm{k}\cdot\bm{r}}u^*_{\alpha\sigma}(\bm{k})c_{\bm{k}\sigma}
\end{equation}
with $\alpha$ being the orbital degrees of freedom. Here the operator $c_{\bm{k}\sigma}$ annihilates a fermion on the isolated band with Bloch wave function $u_{\alpha\sigma}(\bm{k})$.
Meanwhile, the free part $H_0$ is mapped to $H_0=\sum_{\bm{k}\sigma}\varepsilon_{\bm{k}\sigma}c_{\bm{k}\sigma}^\dagger c_{\bm{k}\sigma}$ with $\varepsilon_{\bm{k}\sigma}=\epsilon_{\bm{k}\sigma}-\mu$ being the dispersion spectrum. To deal with the attractive interaction, we introduce the auxiliary Cooper pair operator  $\Delta(\bm{r})=-U a_{-}(\bm{r})a_{+}(\bm{r})$, which gives rise to
 $H_{\mathrm{int}}=\int d\bm{r}\Delta(\bm{r})a^\dagger_{+}(\bm{r})a^\dagger_{-}(\bm{r})+h.c.$. After the Fourier transformation and the band projection, we have 
\begin{equation}
\label{eq:h_interaction}
H_{\mathrm{int}}=\sum_{\bm{k},\bm{q}}\Gamma(\bm{k},\bm{q})\Delta_{\bm{q}}c^\dagger_{\bm{k}+\bm{q},+}c^\dagger_{-\bm{k},-}+h.c..
\end{equation}
Here the form factor $\Gamma(\bm{k},\bm{q})=\sum_{\alpha}u_{\alpha+}(\bm{k}+\bm{q})u_{\alpha-}(-\bm{k})$ as the overlap between Bloch waves, encodes the quantum metric of the Bloch waves which was demonstrated to play a key role in a flat-band superconductors~\cite{peotta2015superfluidity,chen2024ginzburg,hu2025anomalous}.

Generally speaking, one may have $u_{\alpha+}(\bm{k})\neq u^*_{\alpha-}(-\bm{k})$ due to the absence of the $\mathcal{T}$ symmetry. Throughout this work, we focus on the case where the breaking of the $\mathcal{T}$ symmetry is manifested by the polarization of two flavors $\sigma$ (Zeeman effect or valley polarization). For example in the model in Eq.~\eqref{eq:hint}, we assume a valley polarized order $\delta_v$ by changing $\varepsilon_{\bm{k}\sigma}$ to $\varepsilon_{\bm{k}\sigma}+\sigma\delta_v$, which preserves the $U_v(1)$ symmetry~\cite{po2018origin,lee2019theory,bultinck2020mechanism,han2022pair,lin2022zero,hu2023josephson}. 
Thus the wavefunctions maintain as $u_{\alpha+}(\bm{k})= u^*_{\alpha-}(-\bm{k})$. A recent work has studied how $u_{\alpha+}(\bm{k})\neq u^*_{\alpha-}(-\bm{k})$ generates the FFLO phase and the possibility for realizing SDE~\cite{sun2024flat}.

The form factor $\Gamma(\bm{k},\bm{q})$ can be expanded as 
$|\Gamma(\bm{k},\bm{q})|^2=1-\sum_{ab}\mathcal{G}_{ab}(\bm{k})q_a q_b+\mathcal{O}(\bm{q}^3)$ with $\mathcal G_{ab}$ being the quantum metric in Eq.~\eqref{eq:q_matric}. Around the transition region, from the interaction Hamiltonian of Eq.~(\ref{eq:h_interaction}) one may derive the grand potential $F = \int_{\bm{q}}\mathcal F[\Delta_{\bm{q}}]$ with the free energy density up to the fourth order of $|\Delta_{\bm q}|^4$, yielding 
\begin{equation}
\label{eq:free_energy}
\mathcal F[\Delta_{\bm{q}} ]= (a_0+a_{\bm{q}})|\Delta_{\bm{q}}|^2+\frac{b}{2}   |\Delta_{\bm{q}}|^4.
\end{equation}
 The coefficients $a_0$ and $b$ are momentum-independent, which are given by
\begin{equation}
a_0=\frac{1}{U}-T\sum_{\bm{k},n}G_e(\bm{k},\omega_n)G_h(\bm{k},\omega_n)
\end{equation}
and
\begin{equation}
b=T_c\sum_{\bm{k},n}G_e^2(\bm{k},\omega_n)G_h^2(\bm{k},\omega_n).
\end{equation}
Here the Green's functions are $G_e(\bm{k},\omega_n)=(i\omega_n-\varepsilon_{\bm{k},+})^{-1}$ and $G_h(\bm{k},\omega_n)=(-i\omega_n-\varepsilon_{-\bm{k},-})^{-1}$ with $\omega_n=(2n+1)\pi/\beta_c $ being fermionic Matsubara frequencies. We further keep the leading order for $\beta_c W \lesssim 1$ and in the limit $\beta_c W\rightarrow 0$, we have

\begin{align}
a_0&=\frac{1}{W}\tanh \frac{\beta_c W 
\overline T}{4}\label{eq:a0_ginz},\\
b &=\frac{\beta_c}{2W^3}(W-\frac{4}{\beta_c}\tanh \frac{\beta_c W}{4}),
\end{align}
where $\overline T=(T-T_c)/T_c$ is the dimensionless temperature parameter and $T_c=\beta_c^{-1}$ denotes the mean-field transition temperature. 
In particular, $T_c\approx U/4 $ for a narrow band $W\ll U$ at half filling. The $a_{\bm{q}}$ term has the contributions from both the conventional band dispersion and the quantum geometry
$a_{\bm q}= a_{\bm q}^c+a_{\bm q}^g$. The quantum metric contribution vanishes identically in a single-band system, which indicates that $a_{\bm q}^g$ represents the intrinsic multi-band effect.
Due to the absence of $\mathcal{T}$ symmetry, we can have terms with odd orders of $\bm q$ which will account for the SDE. Now we derive the $a_{\bm{q}}^c$ and $a_{\bm{q}}^g$ separately. In doing so, we first note that the second order correction to the free energy is $\delta F_2=1/U-T|\Delta_{\bm{q}}|^2\sum_{\bm{k},n}|\Gamma(\bm{k},\bm{q})|^2 G_{e}(\bm{k}+\bm{q},\omega_n)G_{h}(\bm{k},\omega_n)$. By expanding $\delta F_2$, we can obtain:
\begin{equation}
\label{eq:con_aq}
\begin{split}
a_{\bm{q}}^c=&T_c\sum_{\bm{k},n}[\frac{1}{2}q_a q_b v_a v_b G_h^2 G_e^2\\
&-G_h q_a q_b q_c (\frac{1}{6}v_{abc}G_e^2+v_{ab}v_c G_e^3+v_a v_b v_c G_e^4)]
\end{split}
\end{equation}
and
\begin{equation}
\label{eq:geo_aq}
a_{\bm{q}}^{g}= T_c \sum_{\bm{k},n}[\mathcal{G}_{ab}q_a q_b G_e G_h+q_a q_b q_c \mathcal{G}_{bc} v_aG_e^2 G_h].
\end{equation}
The first term in Eq.~\eqref{eq:con_aq} represents the superfluid density from band dispersion, while the second term quantifies the velocity asymmetry of electrons on the Fermi surface, which is responsible for nonreciprocity. Interestingly, the first term in Eq.~\eqref{eq:geo_aq} describes how the quantum metric contributes to the superfluid density, and the second term corresponds to the quantum metric dipole, which gives rise to band-geometric nonreciprocity. As we will demonstrate below, we extend the general formulas of Eq.~\eqref{eq:con_aq} and Eq.~\eqref{eq:geo_aq} to a specific $C_3$-symmetric system to illustrate both the conventional and geometric origins of the diode effect in a flat-band model Hamiltonian.

To proceed, we consider a system that respects a $C_3$ symmetry and moir\'{e} heterostructures belong to this case. By parametrizing $\bm q=q(\cos\theta,\sin\theta)$ with $\theta$ being the angle between $\bm q$ and the principle axis, we have the following expression up to the $q^3$,
\begin{align}
a_{\bm q}^c&=\frac{\beta_c}{4}\left[ \lambda_c q^2+\delta_v\alpha(\theta)q^3\right].
\label{eq:con_ginz} \\
a_{\bm q}^g&=\frac{\beta_c}{4}\left[\overline{\mathcal{G}}q^2+\delta_v\overline{\mathcal{D}}(\theta) q^3\right],\label{eq:geo_ginz}
\end{align}
where we consider $\delta_v$ as a perturbation $\delta_v\ll W$ and the $C_3$ symmetry ensures that $\overline{\mathcal{G}}_{xx}=\overline{\mathcal{G}}_{yy}\equiv \overline{\mathcal{G}}$ and $\overline{\mathcal{G}}_{xy}=\overline{\mathcal{G}}_{yx}=0$. One may immediately obtain a relation for the effective mass \cite{chen2024ginzburg,hu2025anomalous} of Cooper pairs  $\frac{1}{2m^*}=\beta_c[\lambda_c+\overline{\mathcal{G}}]/4$ with the averaged $\overline{\mathcal{G}}_{ab}$ over the first Brillouin zone, yielding
\begin{align} 
\overline{\mathcal{G}}_{ab}=\sum_{\bm{k}}\mathcal{G}_{ab}(\bm{k})w_{\bm{k}},\label{eq:matric_ave}
\end{align}
with the weight $w_{\bm{k}}=\tanh (\beta_c \varepsilon_{\bm{k}}/2)/(\beta_c \varepsilon_{\bm{k}}/2)$ and $\varepsilon_{\bm k} =\epsilon_{\bm k+}-\mu$. In Eqs.~\eqref{eq:con_ginz} and \eqref{eq:geo_ginz},
the coefficients before $q^3$ depend on the angle $\theta$,
$\alpha(\theta)=\alpha_x^{xx}\cos3\theta-\alpha_y^{yy}\sin3\theta$ and $\overline{\mathcal{D}}(\theta)=\overline{\mathcal{D}}_x^{xx}\cos3\theta-\overline{\mathcal{D}}_y^{yy}\sin3\theta$, which are defined in polar coordinate. Here 
$\alpha_a^{bc}$  is a tensor for Fermi surface asymmetry and $\overline{\mathcal{D}}_a^{bc}$ is the weighted averaged quantum metric dipole with  
\begin{align}
    \overline{\mathcal{D}}_a^{bc} = -\beta_c \sum_{\bm{k}}\partial_a[\mathcal{G}_{bc}(\bm{k})w_{\bm{k}}] f_{\varepsilon_{\bm{k}}},\label{eq:dipole_ave}
\end{align}
with the Fermi-Dirac distribution $f_{\varepsilon_{\bm{k}}}=1/(1+e^{\beta_c \varepsilon_{\bm{k}}})$. By breaking inversion $\mathcal{P}$ symmetry, $\mathcal{G}(\bm{k})\neq\mathcal{G}(-\bm{k})$ leads to $\overline{\mathcal{D}}_a^{bc}\neq0$ as illustrated in Figs.~\ref{fig:fig1}~(a) and (b). In the flat-band limit of $W \ll U$, we have $a_0\rightarrow \overline{T}/4T_c$, $b\rightarrow\beta_c^3/96$, $\overline{\mathcal{G}}\rightarrow \sum_{\bm{k}}\mathcal{G}_{ab}(\bm{k})$ and $\overline{\mathcal{D}}_{a}^{bc}\rightarrow \beta_c^2/4 \sum_{\bm{k}}\mathcal{D}_a^{bc}(\bm{k})\varepsilon_{\bm{k}}$, which is related to Eq.~\eqref{eq:q_dipole}. Besides the geometric terms, the conventional terms in $a_{\bm{q}}^c$ can be obtained as
\begin{equation}
\label{eq:con_lambda}
\lambda_c = \frac{2}{\beta_c^2} \sum_{\bm{k},n}v_x^2 G_e^2 G_h^2,
\end{equation}
and
\begin{equation}
\label{eq:con_diode}
\begin{split}
\alpha_a^{bc} &= -\frac{4}{\beta_c^2} \sum_{\bm{k},n} [\frac{1}{6}v_{abc}(2G_e^3G_h-G_e^2G_h^2)+v_{ab}v_c\\
& (3G_e^4G_h-G_e^3G_h^2)+v_a v_b v_c (4G_e^5G_h-G_e^4G_h^2)],
\end{split}
\end{equation}
where $v_a=\partial_a \varepsilon_{\bm{k}}$, $v_{ab}=\partial_{ab} \varepsilon_{\bm{k}}$ and $v_{abc}=\partial_{abc} \varepsilon_{\bm{k}}$. 

\begin{figure}
		\centering
		\includegraphics[width=1.0\linewidth]{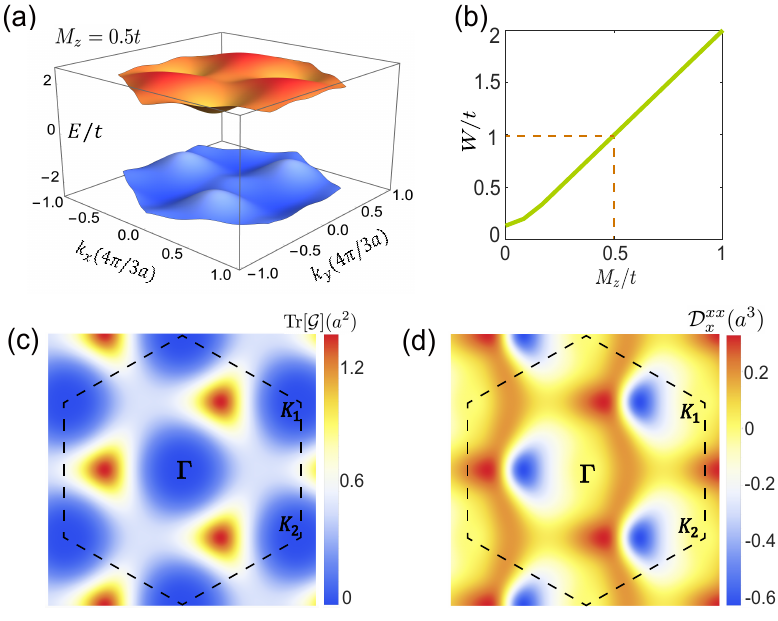}
		\caption{(a) The band structure of the flattened Haldane model with the staggered potential term $M_z=0.5t$. The bandwidth of the lower valence band is approximately $W\approx t$. (b) The bandwidth $W$ as a function of $M_z$. (c), (d) Momentum-space quantum metric $\mathrm{Tr}[\mathcal{G}]$ and quantum metric dipole $\mathcal{D}_x^{xx}$ of the lower band within the first Brillouin zone, which are evaluated from Eq.~\eqref{eq:q_matric} and \eqref{eq:q_dipole}.  In the calculation, we take $t_A^+=t_2e^{i\pi/2}$, $t_B^+=t_2e^{-i\pi/2}$, $t_2=0.39t$ and $t_3=-0.34 t$. Here we denote $t$ as the energy unit.}
		\label{fig:fig2}
\end{figure}

Phenomenologically, the supercurrent $J_{\bm{q}}$ can be derived as $J_{\bm{q}}=2\partial \mathcal{F}_{\bm{q}}/\partial {\bm{q}}$, which gives $J_{\bm{q}}=-2/b(a_0+a_{\bm{q}})\partial_{\bm{q}}a_{\bm{q}}$. $|\Delta_{\bm{q}}|$ has been obtained by minimizing $\mathcal{F}_{\bm{q}}$ as $\partial \mathcal{F}_{\bm{q}}/\partial |\Delta_{\bm{q}}|=0$. With $\delta_v$ as a perturbation ($\delta_v\ll W$), the critical current occurs at the momentum $q_c\approx \pm (-4a_0/[3\beta_c(\overline{\mathcal{G}}+\lambda_c)])^{1/2}$. As a result, we obtain the diode qualify factor $\eta=(J_{c+} - |J_{c-}|)/(J_{c+} + |J_{c-}|)$ as
\begin{equation}
\eta =\frac{\sqrt{3}}{3}|{\bar T}|^{1/2}(\overline{\mathcal{G}}+\lambda_c)^{-\frac{3}{2}}[\overline{\mathcal{D}}(\theta)+\alpha(\theta)]\delta_v.
\label{eq:diode_eta}
\end{equation}
The expression for $\eta$ scales as $\sqrt{T_c - T}$ in the vicinity of the transition temperature. Eq.~\eqref{eq:diode_eta} encapsulates the primary finding of our study. It is evident from Eq.~\eqref{eq:diode_eta} that, in addition to the conventional terms $\lambda_c$ and $\alpha$, the expression for $\eta$ incorporates the quantum metric $\overline{\mathcal{G}}$ and the quantum metric dipole $\overline{\mathcal{D}}$. This insight underscores the significance of quantum geometric factors in determining SDE. 

In the flat-band limit, where $U \gg W$, it follows that $\alpha_a^{bc} = \beta_c^2/24 \sum_{\bm{k}} v_{abc} + \mathcal{O}(\beta_c^4)$ and $\lambda_c=\beta_c^2/24\sum_{\bm{k}}v_{x}^2+ \mathcal{O}(\beta_c^4)$, which are suppressed at the order of $\mathcal{O}(\beta_c^4)$ and $\mathcal{O}(\beta_c^2)$. This results in $\alpha/\overline{\mathcal{D}} \sim W^2/U^2$ and $a_{\bm{q}}^c/a_{\bm{q}}^g \sim W^2/U^2$, indicating that the supercurrent predominantly arises from quantum geometry. As a result, the quantum metric dipole term also dominates the diode effect in this limit. Fig.~\ref{fig:fig1}(d) provides a schematic depiction of the nonreciprocal current-$q$ relation in the presence of $\overline{\mathcal{D}}$. Conversely, when the isolated band becomes highly dispersive, as $\beta_c W \gg 1$, the system transitions back to a conventional BCS regime. In this regime, quantum geometry plays a less significant role, as the averages of $\mathcal{G}$ and $\mathcal{D}$ involve weighted averages near the Fermi surface, which contribute minimally in the conventional scenario, as indicated in Eqs.\eqref{eq:matric_ave} and~\eqref{eq:dipole_ave}.

\begin{figure}
		\centering
		\includegraphics[width=1.0\linewidth]{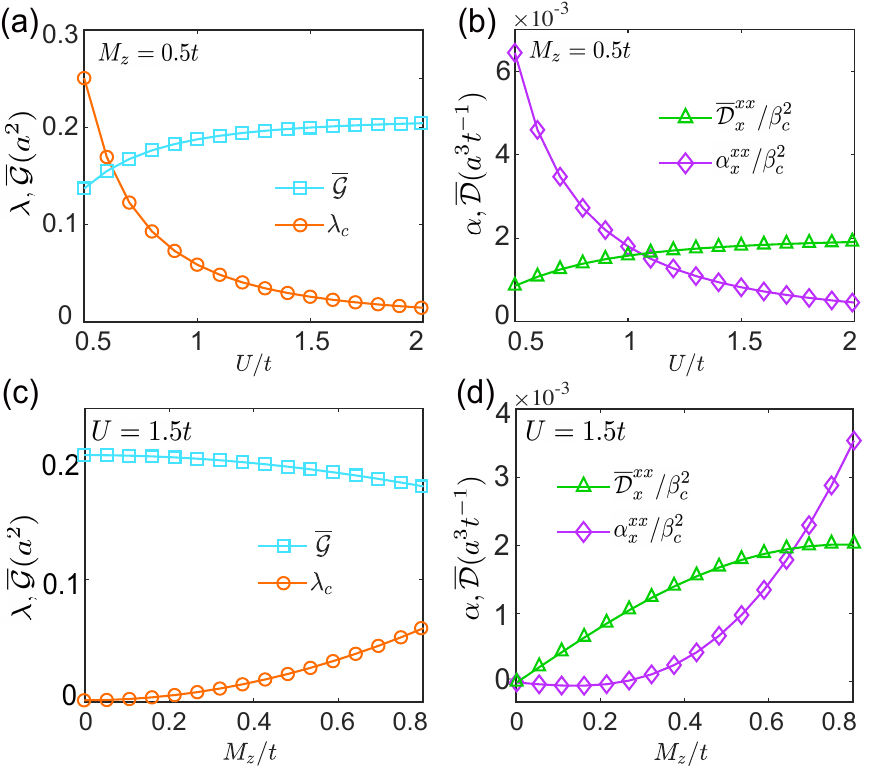}
		\caption{The calculations of the Ginzburg-Landau coefficients for (a) $\lambda_c$, $\overline{\mathcal{G}}$ and (b) $\alpha_x^{xx}$, $\overline{\mathcal{D}}_x^{xx}$ as a function of the attractive interaction $U$ at $M_z=0.5t$. At large $U$, $\overline{\mathcal{G}} \gg \lambda_c$ and $\overline{\mathcal{D}}_x^{xx}\gg \alpha_x^{xx}$. (c) $\lambda_c$, $\overline{\mathcal{G}}$ and (d) $\alpha_x^{xx}$, $\overline{\mathcal{D}}_x^{xx}$ as a function of $M_z$ at $U=1.5t$. In (d) $\alpha_x^{xx}=\overline{\mathcal{D}}_x^{xx}=0$ when $M_z=0$. $M_z$ is the staggered potential in Eq.~(\ref{eq:Haldane_lattice}) which breaks the $\mathcal{P}$ symmetry. The Fermi energy is at $\mu=-2t$.}
		\label{fig:fig3}
\end{figure}
Several key points can be extracted from the Ginzburg-Landau framework: (i) The nonreciprocity of the supercurrent arises from its part Fermi surface asymmetry $\alpha$ and part quantum metric dipole $\overline{\mathcal{D}}$. (ii) Although the conventional and geometric supercurrents coexist in genenal cases, in the flat-band limit when $U\gg W$, the geometric component predominates over the conventional one. (iii) It is crucial to highlight that in the exact flat-band limit with $W=0$, $\lambda_c=\alpha=\overline{\mathcal{D}}=0$, and only $\overline{\mathcal{G}}$ is nonzero, leading to purely geometric supercurrent but no diode effect. In the following section, we will demonstrate these behaviors by calculating the Ginzburg-Landau coefficients for a Haldane-like flat-band model. Furthermore, for realistic materials such as moir\'{e} graphene, the fundamental physics of the low-energy flat bands can be effectively described by a Haldane-like model~\cite{zhang2019twisted}.

\begin{figure*}
		\centering
		\includegraphics[width=1.0\linewidth]{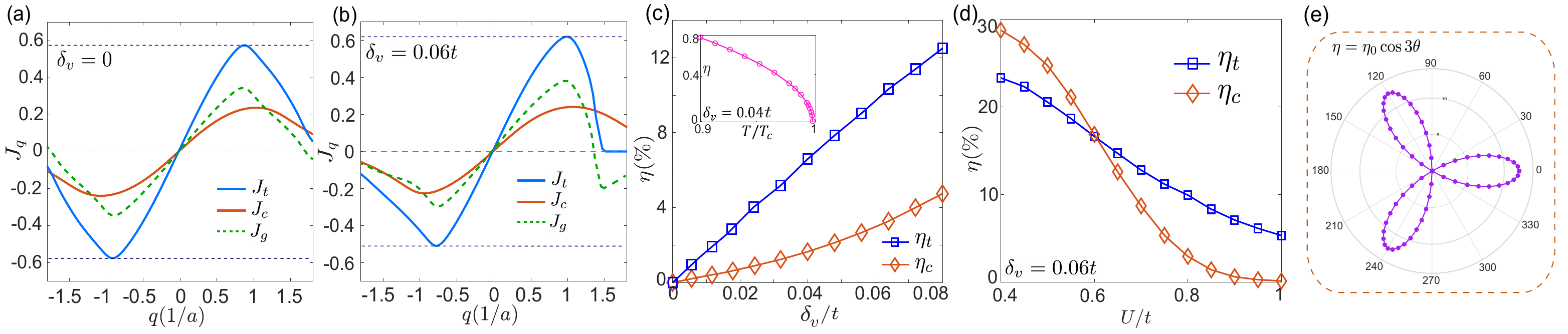}
		\caption{(a) The $q$ dependence of the supercurrent $J_q$ without valley polarization $\delta_v$. The total and conventional currents, denoted as $J_t$ and $J_c$, are plotted in blue and orange lines, respectively. The current from quantum geometry ($J_g$) is plotted in the green dashed line. In this case, there is no diode effect. (b) The $q$ dependence of the $J_q$ at $\delta_v=0.06t$. In this case, $J_t$, $J_c$ and $J_g$ all exhibit nonreciprocal behaviors. (c) The diode quality factors of total current $\eta_t$ and conventional current $\eta_c$ as a function of $\delta_v$. The inset shows the temperature dependence of $\eta_t$ near $T_c$ at $\delta_v=0.04t$. (e) $\eta_c$ and $\eta_t$ as a function of $U$. (e) The angular dependence of $\eta$ with $\eta=\eta_0 \cos 3\theta$ and $\eta_0=12\%$ at $\delta_v=0.08t$ and $U=0.8t$. Parameters: for all five panels, $\mu=-2.15t$, $T=0.02t$, $M_z=0.5t$.}
		\label{fig:fig4}
\end{figure*}
\section{Flat-band model study}
To illustrate the features of the Ginzburg-Landau theory for SDE developed above, we examine a Haldane-like model with two valleys and two orbitals ($A$ and $B$) per-site on the triangular lattice \cite{yang2012topological}, which has the form in real space
\begin{equation}
\label{eq:Haldane_lattice}
\begin{split}
h_0^\sigma=&-\sum_{\braket{ij}}t a_{i\sigma}^{\dagger}b_{j\sigma}-\sum_{\braket{ij}'}t_A^\sigma a_{i\sigma}^{\dagger}a_{j\sigma}+t_B^\sigma b_{i\sigma}^{\dagger}b_{j\sigma}\\
&-\sum_{\braket{ij}''}t_3 a_{i\sigma}^{\dagger}b_{j\sigma}+\sum_i M_z(a_{i\sigma}^{\dagger}a_{i\sigma}-b_{i\sigma}^{\dagger}b_{i\sigma}),
\end{split}
\end{equation}
where $\braket{ij}$, $\braket{ij}'$ and $\braket{ij}''$ represent the first, second and third nearest neighbor hoppings, respectively. $\sigma=\pm 1$ denotes the valley, and operator $a(b)_{i\sigma}$ annihilates a fermion with $\sigma$ valley in the orbital $A$($B$) on site $i$. $M_z$ is the staggered potential term on the $A$ and $B$ orbitals. Due to $\mathcal{T}$ symmetry, $t_{A/B}^-=(t_{A/B}^+)^*$. The energy spectrum of the model is shown in Fig.~\ref{fig:fig2}(a), which exhibits a pair of narrow bands at $E\approx \pm 2t$. It is straightforward to obtain the momentum-space distribution of the quantum metric and quantum metric dipole for the lower band at $M_z=0.5t$ from Eq.~\eqref{eq:q_matric} and \eqref{eq:q_dipole} as plotted in Fig.~\ref{fig:fig2}(c) and (d). In Fig.~\ref{fig:fig2}(b), increasing $M_z$ enhances the bandwidth $W$. Evidently, $M_z$ breaks $\mathcal{P}$ symmetry, leading to $\mathcal{G}_{ab}(\bm{k})\neq \mathcal{G}_{ab}(-\bm{k})$ and $\mathcal{D}_{a}^{bc}(\bm{k})\neq -\mathcal{D}_{a}^{bc}(-\bm{k})$.

We calculate the Ginzburg-Landau coefficients from Eq.~(\ref{eq:matric_ave}) to Eq.~(\ref{eq:con_diode}), where the Fermi energy resides on the lower band ($\mu=-2t$), as depicted in Fig.~\ref{fig:fig2}(a). Fig.~\ref{fig:fig3}(a) displays the quadratic terms $\lambda_c$ and $\overline{\mathcal{G}}$, plotted as functions of $U$. As expected, $\overline{\mathcal{G}}$ reaches its optimal value at high values of $U$, while $\lambda_c$ exhibits a power-law decay, scaling as $1/U^2$. Consequently, the supercurrent contribution from the quantum metric becomes predominant over the conventional component in the regime of large $U$. In Fig.~\ref{fig:fig3}(b), we illustrate the cubic terms $\alpha_{x}^{xx}$ and $\overline{\mathcal{D}}_x^{xx}$, which decay as $1/U^2$ and $1/U^4$, respectively. Clearly, the nonreciprocal term, driven by the quantum metric dipole, overshadows the conventional Fermi surface asymmetry term $\alpha_{x}^{xx}$. Furthermore, the staggered potential term $M_z$, which violates $\mathcal{P}$ symmetry, plays a critical role in the SDE. This is shown Fig.~\ref{fig:fig3}(c) and (d), where an increase in $M_z$ causes $\lambda_c$ to rise from nearly zero due to bandwidth enhancement, while $\overline{\mathcal{G}}$ remains almost unaffected. Fig.~\ref{fig:fig3}(d) shows that both $\overline{\mathcal{D}}_{x}^{xx}$ and $\alpha_x^{xx}$ vanish exactly when $M_z = 0$, yet display different trends as $M_z$ increases. This behavior aligns with the understanding of Eq.~(\ref{eq:con_diode}) due to the enhanced band velocity by $M_z$.

\section{Mean-field calculations for SDE}
The Ginzburg-Landau analysis provides a phenomenological perspective near $T_c$ and within the long-wavelength regime ($q\ll 1/a$, $a$ is the lattice constant). To study the supercurrent within a microscopic framework that takes into account quantum geometry, we would like to present a mean-field study for the lattice model introduced in Eq.~\eqref{eq:Haldane_lattice}. Also for the realistic material such as moir\'{e} graphene, the main physics of the low-energy flat bands can be captured by a Haldane-like model~\cite{zhang2019twisted}.

By including the pairing potential $\Delta_{\bm{q}}$, the Bogoliubov-de Gennes (BdG) Hamiltonian in the Nambu basis $\Psi_{\bm{k},\bm{q}}=(c_{\bm{k}+\bm{q},+},c^\dagger_{-\bm{k},-})^T$ reads
\begin{equation}
H_{\mathrm{BdG}}^{\bm{k},\bm{q}}=\left(
\begin{matrix}{}
  &\epsilon_{\bm k+\bm{q},+}-\mu+\delta_v & \Delta_{\bm{q}}\Gamma(\bm{k},\bm{q})  \\
  &\Delta_{\bm{q}}\Gamma^*(\bm{k},\bm{q})  &  -\epsilon_{-\bm k,-}+\mu+\delta_v
\end{matrix}\right).
\end{equation}
Thus the total free energy can be written as 
\begin{equation}
F(\bm{q})=-\frac{1}{2\beta}\sum_{\bm{k},n}\mathrm{ln}[1+e^{-\beta E_n(\bm{k},\bm{q})}]+\frac{|\Delta_{\bm{q}}|^2}{2U}
\end{equation}
Here $E_n(\bm{k},\bm{q})$ is the eigenvalues of $H_{\mathrm{BdG}}$. $\Delta_{\bm{q}}$ need to be determined self-consistently and the total supercurrent can be derived as $J_{\bm{q}}=2\partial_{\bm{q}}F(\bm{q})$. The conventional supercurrent can be obtained with $\Gamma=1$.

We adopt the band-projection method to study the supercurrent self-consistently where the Fermi energy lies on the lower band in Fig.~\ref{fig:fig2}~(a). The supercurrent from conventional band dispersion ($J_c$) and the total one ($J_t$) incorporated with the quantum geometry can be computed directly, while the geometric contribution is approximately $J_g\approx J_t-J_c$. Along the $x$ direction with $\bm{q}=q \hat{x}$, the supercurrent as a function of $q$ is shown in Fig.~\ref{fig:fig4}~(a) and (b), in the absence and presence of the valley polarization $\delta_v$, respectively. In Fig.~\ref{fig:fig4}(a) there is no diode effect because $\delta_v=0$ to preserve the $\mathcal{T}$ symmetry. In Fig.~\ref{fig:fig4}(b), the diode effect shows up for finite $\delta_v$. Obviously, besides the $J_c$ and $J_t$, the $J_g$ is also nonreciprocal arising from the quantum metric dipole. In Fig.~\ref{fig:fig4}(c) we find that the diode quality factor $\eta$ grows linearly for small $\delta_v$ in both the total ($\eta_t$) and conventional ($\eta_c$) cases. The inset reveals the temperature scaling of $\eta$ as $\sqrt{T_c-T}$ near $T_c$. In Fig.~\ref{fig:fig4}(d) we show the $U$ dependence of $\eta_c$ and $\eta_t$. As $U$ increases, $\eta_c$ decays faster than $\eta_t$ and $\eta_t\gg \eta_c$ when $U$ is large, indicating that the quantum geometry dominates in the large $U$ regime. Since the geometric supercurrent can not be calculated directly, it can be inferred that the quantum metric dipole plays a pivotal role for the SDE in the large $U$ regime, which is consistent with the Ginzburg-Landau analysis. We also show the angular dependence of $\eta$ in Fig.~\ref{fig:fig4}(e) by using $\bm{q}=q(\cos\theta, \sin\theta)$ and verify that $\eta \sim \cos(3\theta)$.

\section{Conclusion}
In summary, we have formulated the Ginzburg-Landau analysis and mean-field  calculations for the SDE in narrow-band superconductors, with particular applicability to moir\'{e} graphene systems. Our work highlights the essential contributions of the quantum metric and quantum metric dipole to the non-dissipative transport and resulting in SDE. Unfortunately, considering the weak-coupling regime in most of superconductors in reality, the conventional contribution from band dispersion effect may still dominate in the SDE. It is still an open question that if a system can have SDE dominated by quantum geometry. Interestingly, the quantum metric dipole plays a vital role in other nonreciprocal phenomena~\cite{gao2019nonreciprocal,wang2021intrinsic,liu2021intrinsic,arora2022quantum,kaplan2024unification}. Although our phenomenological theory primarily focuses on $C_3$ symmetric systems with a specific way for breaking $\mathcal{T}$, it is highly general and the method can be extended to other cases as well.

\section{Acknowledgements}
We appreciate inspiring discussions with Jiang-Xiazi Lin, Bo Yang and Justin C. W. Song. K.T.L. acknowledges the support of the Ministry of Science and Technology, China, and Hong Kong Research Grant Council through Grants No. 2020YFA0309600, No. RFS2021-6S03, No. C6053-23G, No. AoE/P-701/20, No. 16310520, No. 16310219, No. 16307622, and No. 16311424.

	\appendix
	\renewcommand{\theequation}{A-\arabic{equation}}
	\renewcommand\thefigure{A-\arabic{figure}}
	\setcounter{equation}{0}
	\setcounter{figure}{0}
	
\section{Ginzburg–Landau functional for SDE}\label{AppendixA}
In this section we develop the Ginzburg Landau theory for the SDE and consider the quantum geometry effects. We use the Green's function approach while the same result can also be obtained by path integral~\cite{chen2024ginzburg}. We can start from the interaction Hamiltonian~\eqref{eq:h_interaction} as
\begin{equation}
H_{\mathrm{int}}=\sum_{\bm{k}}\Gamma(\bm{k},\bm{q})\Delta_{\bm{q}}c^\dagger_{\bm{k}+\bm{q},+}c^\dagger_{-\bm{k},-}
\end{equation}
with the form factor $\Gamma(\bm{k},\bm{q})=\sum_{\alpha}u_{\alpha+}(\bm{k}+\bm{q})u_{\alpha-}(-\bm{k})$. From the standard perturbation theory, we can obtain the second-order and fourth-order Ginzburg-Landau coefficients as
\begin{align}
\delta F_2 &=\frac{1}{U}-T|\Delta_{\bm{q}}|^2\sum_{\bm{k},n}|\Gamma(\bm{k},\bm{q})|^2 G_{e}(\bm{k}+\bm{q},\omega_n)G_{h}(\bm{k},\omega_n) \\
\delta F_4&=\frac{T_c}{2}|\Delta_{\bm{q}}|^4\sum_{\bm{k},n}G_{e}(\bm{k},\omega_n)^2 G_{h}(\bm{k},\omega_n)^2.
\end{align}
The total free energy in the Eq.(6) of main text is written as
\begin{equation}
\Omega=\int d\bm{q} (\delta F_2+\delta F_4)=\int d\bm{q} [(a_0+a_{\bm{q}})|\Delta_{\bm{q}}|^2+\frac{b}{2}|\Delta_{\bm{q}}|^4]
\end{equation}
Here the $a_{\bm{q}}=a_{\bm{q}}^c+a_{\bm{q}}^g$ includes both the conventional and geometric parts. We first derive $a_{\bm{q}}^c$, which reads
\begin{equation}
\begin{split}
&a_{\bm{q}}^c=-T_c\sum_{\bm{k},n}[G_e(\bm{k}+\bm{q},\omega_n)G_h(\bm{k},\omega_n)-G_e(\bm{k},\omega_n)G_h(\bm{k},\omega_n)]\\
&\approx -T_c\sum_{\bm{k},n}G_h[\bm{q}\cdot \nabla_{\bm{k}}+\frac{1}{2}(\bm{q}\cdot \nabla_{\bm{k}})^2+\frac{1}{6}(\bm{q}\cdot \nabla_{\bm{k}})^3]G_e
\end{split}
\end{equation}
Here we use the notations that $G_e=G_e(\bm{k},\omega_n)$ and $G_h=G_h(\bm{k},\omega_n)$. $\nabla_{\bm{k}}$ is the derivative of the wave number. The first-order term is zero due to the rotation symmetry. This gives Eq.~\eqref{eq:con_diode} of main text. The geometric term $a_{\bm{q}}^{g}$ reads
\begin{equation}
a_{\bm{q}}^{g}=T_c \sum_{\bm{k},n} (1-|\Gamma(\bm{k},\bm{q})|^2)G_e(\bm{k}+\bm{q},\omega_n)G_h(\bm{k},\omega_n).
\end{equation}
Here the $|\Gamma(\bm{k},\bm{q})|$ is the form factor and can be expanded as
\begin{equation}
|\Gamma(\bm{k},\bm{q})|^2=|\langle u_{\bm{k}+\bm{q}}|u_{\bm{k}}\rangle|^2=1-\mathcal{G}_{ab}q_a q_b+ \mathcal{O}(\bm{q}^3)
\end{equation}
Based on this, we can derive that 
\begin{equation}
\begin{split}
a_{\bm{q}}^{g}&=T_c \sum_{\bm{k},n} \mathcal{G}_{ab}q_a q_b G_e(\bm{k}+\bm{q},\omega_n)G_h(\bm{k},\omega_n)\\
&= T_c \sum_{\bm{k},n}\mathcal{G}_{ab}q_a q_b[G_e(\bm{k},\omega_n)G_h(\bm{k},\omega_n)\\
&+q_c v_c(\bm{k})G_e^2(\bm{k},\omega_n)G_h(\bm{k},\omega_n)]
\end{split}
\end{equation}

For the specific $C_3$-symmetric system, we can expanding them and write them as $a_{\bm{q}}^c=\frac{\beta_c}{4}(\lambda_c \bm{q}^2+\delta_v\alpha_a^{bc}q_a q_b q_c)$ and $a_{\bm{q}}^g=\frac{\beta_c}{4}(\overline{\mathcal{G}}_{ab} q_a q_b+\delta_v\overline{\mathcal{D}}_a^{bc}q_a q_b q_c)$. $\lambda_c$ is the conventional kinetic energy term which is related to the Fermi velocity $v_F$. $\alpha_a^{bc}$ is the tensor which describes the Fermi surface asymmetry such as trigonal warping. $\overline{\mathcal{G}}_{ab}$ and $\overline{\mathcal{D}}_a^{bc}$ are the average of quantum metric and quantum metric dipole which are defined in the main text. Because of the rotation symmetry, we have $\overline{\mathcal{G}}_{xx}=\overline{\mathcal{G}}_{yy}\equiv\overline{\mathcal{G}}$ and $\overline{\mathcal{G}}_{xy}=\overline{\mathcal{G}}_{yx}=0$. Thus we can write down the $a_{\bm{q}}$ term in the Ginzburg-Landau free energy as
\begin{equation}
a_{\bm{q}}=\frac{\beta_c}{4}[(\lambda_c+\overline{\mathcal{G}})\bm{q}^2+\delta_v(\alpha_a^{bc}+\overline{\mathcal{D}}_a^{bc})q_a q_b q_c]
\end{equation}
We analyse the strong coupling limit of $\alpha_a^{bc}$ and $\lambda_c$. In the limit of $\beta_c W \rightarrow 0$, $\sum_n 2G_e^3G_h-G_e^2G_h^2 =-\frac{1}{16}\beta_c^4+\mathcal{O}(\beta_c^6)$, $\sum_n 3G_e^4G_h-G_e^3G_h^2 =\frac{1}{48}\varepsilon_{\bm{k}}\beta_c^6+\mathcal{O}(\beta_c^8)$, and $\sum_n 4G_e^5 G_h-G_e^4G_h^2 =\frac{1}{96}\beta_c^6+\mathcal{O}(\beta_c^8)$. By keeping the leading order, $\alpha_{a}^{bc}\rightarrow \frac{\beta_c^2}{24}\sum_{\bm{k}}v_{abc}+\mathcal{O}(\beta_c^4)$ in the limit of $\beta_c W\rightarrow 0$. With the periodic boundary condition, $\sum_{\bm{k}}v_{abc}=0$ and $\alpha_{a}^{bc}$ scales as $\beta_c^4$. We also have $\lambda_c\sim (\beta_c W)^2$. 

Because of $C_3$ symmetry, $a_{\bm{q}}$ should be invariant by changing $\bm{q}\rightarrow \mathcal{R}_{2\pi/3}\bm{q}$ with $\mathcal{R}$ being the rotation matrix. We further define $\mathcal{C}_a^{bc}=\alpha_a^{bc}+\overline{\mathcal{D}}_a^{bc}$. With respect to the $C_3$ symmetry, we have
\begin{eqnarray}
\overline{\mathcal{C}}_x^{yy}=\overline{\mathcal{C}}_y^{yx}=\overline{\mathcal{C}}_y^{xy}=-\overline{\mathcal{C}}_x^{xx}\\
\overline{\mathcal{C}}_y^{xx}=\overline{\mathcal{C}}_x^{xy}=\overline{\mathcal{C}}_x^{yx}=-\overline{\mathcal{C}}_y^{yy}
\end{eqnarray}
Thus the non-equivalent non-zero elements are $\overline{\mathcal{C}}_x^{xx}$ and $\overline{\mathcal{C}}_y^{yy}$. We have $\sum_{abc}\overline{\mathcal{C}}_a^{bc}q_a q_b q_c=\overline{\mathcal{C}}_x^{xx}(q_x^3-3q_x q_y^2)+\overline{\mathcal{C}}_y^{yy}(q_y^3-3q_y q_x^2)=\overline{\mathcal{C}}_x^{xx}\cos 3\theta-\overline{\mathcal{C}}_y^{yy}\sin 3\theta$.

\end{document}